\documentstyle[preprint,aps]{revtex}
\def\hts{high temperature superconductors}

\def\HTY{High Temperature Superconductivity}
\begin{document}
\flushbottom
\draft
\title{
Topological Doping of Correlated Insulators}
\author{S.~A. ~Kivelson}
\address{
Dept. of Physics,
UCLA,
Los Angeles, CA  90024}
\author{V.~J. ~Emery}
\address{
Dept. of Physics,
Brookhaven National Laboratory,
Upton, NY  11973}
\maketitle
 
\begin{abstract}  

A material which is an insulator entirely because of interaction effects is called
a correlated insulator.  Examples are trans-polyacetylene and the cuprate high 
temperature superconductors.  Whereas doping of a band insulator results in a shift
of the chemical potential into the conduction or valence band, doping of a correlated
insulator produces fundamental changes in the electronic density of states itself.  We
have found that a general feature of doping a correlated insulator is  the generation
of topological defects;  solitons in one-dimension and anti-phase domain walls in
higher dimensions.  
We review the well known features of this process in polyacetylene, and
describe the experimental evidence that the analogous features are seen in 
the cuprate superconductors.  We also distinguish the case in which the doping-induced
features can be viewed as a Fermi surface instability, as in
polyacetylene, and the more usual case
in which they are a consequence of a Coulomb frustrated electronic tendency
to phase separation.

\noindent{\it  Paper presented at the 
Sixtieth Birthday Symposium for Alan J. Heeger, January 20, 1996}

\end{abstract}

\section{Introduction}

Perhaps the most intensively studied problem in condensed matter physics
in recent years is the doping of a Mott insulator,
a system with one electron per unit cell, which is insulating
rather than metallic by virtue of the strong electron-electron repulsion.  
This long-standing unsolved problem has been revitalized by
the discovery, almost a decade ago, of the cuprate high temperature 
superconductors, which are themselves doped Mott insulators.
The scale of temperatures below which a system
is insulating is set by the gap in the charge excitation spectrum; 
the interactions are sufficiently 
strong that this scale is much larger than the temperature of any transition to an 
electronically-ordered state. Nonetheless, at least on a bipartite lattice,
such systems typically exhibit ordered ground states in which the 
size of the unit cell doubles.  Of course, the most-studied example 
has a ground state with N{\'e}el, or  antiferromagnetic order, 
as in the high temperature superconductors.

This problem is a subclass of the somewhat broader problem of 
doping a correlated insulator, which is relevant  for a host
of synthetic metals that have been studied in the past three decades.
By a correlated insulator, we simply mean a material that would be a metal
in a single-electron theory, but is rendered insulating
by interaction effects\cite{elphonon}. As with the Mott insulator, the commensurability
between the density of conduction electrons and the underlying 
lattice is important. Frequently the unit cell contains one conduction 
electron or hole in the high-temperature phase, and is doubled in the broken 
symmetry state.

It is our contention that a certain more or less universal motif,
which we have named ``topological doping'', characterizes the
evolution of the electronic structure of such materials. 
The basic idea is that it is energetically favorable for the system to 
maintain the local commensurate insulating structure even when lightly doped,
and for the additional charges to be incorporated as local defects. 
Moveover we find, quite generally, that the preferred defects
are topological, in that the commensurate structure experiences a phase
shift across the defect. When there is one conduction electron or 
hole  per unit cell, the phase shift is equal to $\pi$.
In one dimension such defects are $\pi$ solitons,
while in higher dimensions they are the various forms of discommensuration,  
familiar from studies of the commensurate-incommensurate transition\cite{pro}.
It is a natural consequence of this view that, as the doping is increased, 
there is a crossover from a ``doped insulator'' regime, where the
defects are far separated, to an ``overdoped'' or more or less conventionally
metallic regime, when the spacing between defects is comparable to their width\cite{em}.

Many materials display the characteristics of topological doping, and it is 
useful to take a broad perspective in order to understand what they have in common.
In this paper, we use the language of topological doping to describe
the well known polyacetylene story, whose understanding was largely based
on the work of A. J. Heeger and collaborators\cite{review}.  We then show that
analagous considerations govern the evolution
of the electronic structure of the high temperature superconductors.  

\section{Trans-polyacetylene}

{\it The Undoped State:}  Undoped trans-polyacetylene\cite{review}
has a half-filled $\pi$ band on the order of 10 eV wide,
and it would be a rather good metal were it not for the Peierls instability
that causes it to dimerize;  thus it is a prime example of a correlated
insulator.  The dimerized state is insulating and characterized
by a broken symmetry, {\it i.e.} the dimerization itself, which can be observed
as a Bragg peak in X-ray scattering.  (The existence and magnitude of the
static dimerization also has been inferred from fancy NMR techniques.)
The undoped polymer exhibits a gap
to both charge and spin excitations.  The charge gap can be seen,
for example, in the optical absorption spectrum and
also has been measured by electrochemical means.
The existence of a spin gap can be deduced from the vanishingly small
value of the spin susceptibility\cite{currie}, although 
its magnitude has been obtained only by indirect means, such as
third-harmonic generation spectroscopy\cite{Wu}.  The gap in the 
excitation spectrum of a single electron, which of course combines charge and spin,
also has been measured by photoemission spectroscopy\cite{photoem}.  
All of these measurements are more or less
consistent with a semiconducting model of the insulating state with a
gap of order 1.8 eV.  There are, however, 
two features of the insulating state that suggest the inadequacy
of the semiconductor model:  {\bf i)}  Long sub-gap tails, observed
by for example optical absorption spectroscopy, suggest that there are 
excitations which have a substantially lower energy and are weakly accessible 
in the insulating state.
These tails are now thought to be due to the direct photoproduction of soliton
pairs, and are weak since they rely on the existence of virtual solitons as
quantum fluctuations in the ground state.  {\bf ii)}  There are remarkable
photo-induced changes in the electronic structure as seen, for example,
in photo-absorption experiments.  These changes are
known to be directly analagous to the dopant-induced changes in the electronic
structure, and hence these experiments can be viewed as exploring
the effects of photodoping the polymer.

{\it The Doped Insulator:}
Upon doping,   the conductivity of  trans-polyacetylene rises by
many orders of magnitude, signifying that there is no longer a gap
(or at most a very small gap) in the charge
excitation spectrum. This can be seen in more detail from
the dopant-induced changes in the optical absorption
spectrum.  Here, it is found that, upon doping, the spectral weight of
the features associated with the insulating gap is transferred 
continuously (and rapidly) to a broad ``mid-gap'' absorption feature, with 
substantial weight extending to much lower frequencies.  At the same time,
the spin susceptibility remains very small, suggesting
that the gap in the spin excitation spectrum is not destroyed.  
As far as we know, the
symmetry of the ground state is restored even at the lowest doping levels\cite{symm}.

The basic explanation of these features was set forward in the seminal paper by
W-P. Su, J. R. Schrieffer, and A. J. Heeger in 1979\cite{ssh}.  They realized
that doping does not proceed uniformly, as in a semiconductor, but rather
it is energetically favorable to destroy the dimerization, and hence the gap, in local 
``defect'' regions,
associated with antiphase domain walls  or discommensurations in the dimerization
order.  These are the famous solitons in polyacetylene.
The local nature of the dopant-induced changes in the
dimerization is responsible for most of the salient features of the
doping of polyacetylene:  {\bf 1)} The spin gap remains intact because the
dimerization (hence the gap) persists between solitons.
{\bf 2)} In the core of the soliton, the dimerization vanishes, hence there
exists a zero energy (midgap) electronic bound state associated with each 
soliton. This mid-gap state leads to the shift of density of states into
the middle of the insulating gap and hence to the associated peak in the 
optical absoption, with weight
proportional to the dopant concentration.  (The fact that all bound states are
either fully occupied or unoccupied implies that there are no low-energy spin
excitations associated with the soliton cores.)
{\bf 3)}  Low energy collective charge excitations are 
associated with the translational motion of the domain walls 
and appear in the far infrared absorption spectrum with
oscillator strength proportional to the dopant concentration.

It is a more subtle matter to deduce the topological character of 
the dopant-induced defects
directly from experiments.  
Soliton doping (as opposed to, for example, bipolaron
doping) produces different distributions of states in the gap, although this is
complicated by interaction effects and disorder.  But, most importantly, the
rapid restoration of the symmetry of the ground state upon doping is a 
consequence of the topological character of the defects;  dopant-induced non-topological
defects would produce only a modest reduction in the magnitude of the order parameter
in proportion to the dopant concentration.  One additional consequence of the topological
character of the solitons is that they are anomalously one-dimensional, since
a single soliton cannot tunnel between two polymer chains.  This is a form
of ``confinement'',
and is a general feature of the one-dimensional electron gas in the spin gap
regime\cite{spingapconfine}. Confinement 
has not been directly observed experimentally, 
but it is fair to say that no evidence of three-dimensional 
quantum mechanical coherence has ever been observed in these
materials, despite the fairly substantial magnitude of the wave function overlap
between adjacent polyacetylene chains.  

Of course, as a consequence of the dynamics of the solitons, the above picture 
of an inhomogeneous dopant-induced state is
valid only at intermediate time scales and length scales.  In the absence of
disorder, and for weak enough interchain coupling, thermal and
quantum flucutations of the soliton positions produce a ``soliton liquid'' 
state\cite{longdist}.  Here, dimerization is completely absent in the
static structure factor, but would be evident at intermediate times
(or frequencies).  In the more realistic case in which disorder
plays a dominant role, the solitons are pinned at random positions, producing
a ``soliton glass'' state\cite{KE}.  In either case, all 
configuration-averaged static correlation functions
ultimately are translationally invariant or, equivalently,
the dimerization order parameter is zero when averaged over distances long 
compared to the mean spacing between solitons.

Finally, we observe that the conductivity of
lightly-doped polyacetylene,
although enormous compared to the undoped polymer, is still considerably
too small to be thought of as metallic in the usual sense, since it
violates the so-called Ioffe-Regel criterion\cite{IR,bad}.  Basically, 
the idea is that the smallest value of the conductivity consistent with Boltzmann
transport theory occurs when the mean-free-path, $\ell$, is equal to
the Fermi wavelength, $\lambda_F=2\pi/k_F$, and this constitutes a sort
of minimum conductivity, $\sigma_{m}$, for a conventional metal. 
Any material with a smaller conductivity
cannot be understood in terms of the occasional scattering of coherently
propagating quasiparticles.  For a quasi one-dimensional material, such as
polyacetylene, $\sigma_{m}=4(e^2/\hbar) 
(n_{\perp}/k_F)\approx 1.5 m\Omega-cm$, where
$n_{\perp}$ is the areal density of polyacetylene chains.  In the entire
doped-insulator regime, the conductivity at room 
temperature is substantially smaller
than $\sigma_{m}$, and it decreases with decreasing temperature.
This poor conductivity is
a consequence of  
soliton doping. The fact that the solitons are confined to one dimension implies that
they are easily pinned by disorder.  This is exacerbated by their
local character, in that the majority of the electrons remain
more or less condensed in the insulating state, and hence cannot participate in
screening the impurity potential. 
If, by contrast, doping supressed all dimerization
entirely, the room temperature conductivity
would be large\cite{KH}, as it is in the overdoped regime.

{\it The Overdoped Metal:}
All of the above properties are features of the lightly-doped polymer which
one can think of as a ``doped insulator''.  Clearly, there is a characteristic
dopant concentration 
$y_c$, which is roughly equal to the reciprocal of the
soliton width, at which the defects begin to overlap.  
For higher dopant concentrations,
one expects a crossover or transition to new physics.  This crossover is
clearly evident in the abrupt destruction of the spin gap above
a critical dopant concentration which is of order $y_c \approx 5\%$,
although it depends on the dopant species.
The overdoped polymer behaves 
grossly like a three-dimensional Fermi liquid\cite{KH}: 
{\bf 1)}  It has a Pauli susceptibility and a T-linear term 
in the specific heat of roughly the
magnitude that would be expected from band theory.  {\bf 2)}  
Its  D.C. conductivity
can be as large as that of typical metals, and certainly
much larger than $\sigma_{m}$.  {\bf 3)}  It exhibits three dimensional
coherent transport, as seen for instance in the measured 
negative magnetoresistance \cite{magnetoresist}.

The essential known features of the topological 
doping of trans polyacetylene are
summarized in Table I.

\section{Coulomb frustrated electronic phase separation}

In a series of previous publications, we have worked out the
theory of topological doping of correlated insulators in
higher dimensions, with special emphasis on 
the high temperature superconductors
\cite{LA,conf,ute,spherical,salkola,others}.
While the ideas are quite general, we shall use language specific to the high 
temperature superconductors in summarizing the basic ideas,
{\it i.e.} we will take the commensurate insulating state
to be an ordered spin 1/2 Heisenberg antiferromagnet, and the doping to be p-type,
{\it i.e.} hole doping:

{\bf 1)}  Given a short-range model of a doped antiferromagnet, such as the $t-J$ model, 
we have shown that the most common behavior at low doping concentrations
is phase separation.  That is to say,
the system separates into two macroscopically distinct regions:  a commensurate
ordered insulating region and a hole-rich metallic region.  It is 
important to recognise that the driving force
for this electronic phase separation is kinetic;  it derives from the fact that
the hole motion is frustrated in the antiferromagnetic state.  Thus, phase
separation is the best compromise between the hole kinetic energy and antiferromagnetic
ordering. Since superexchange is itself a kinetic process, the 
strongly-correlated system has chosen phase separation as a means of minimizing
the total zero-point energy.
Elsewhere we have reviewed the set of model systems which can be proven
to exhibit phase separation, and the experimental evidence that this physics
is operative in the high temperature superconductors\cite{LA}.
Among other things, high temperature superconductors exhibit
macroscopic phase separation whenever the dopants are mobile enough to
counterbalance the effect of the long-range Coulomb
interaction, as in the case of photodoping or oxygen doping. 
In other words ``turning off'' the
long-range Coulomb interaction reveals the microscopic tendency to phase
separation in these materials.
     
{\bf 2)}   In the presence of long-range Coulomb interactions, the
short-range tendency to phase separation
is frustrated.  So long as the
Coulomb interaction is not too strong, this always results in a state which
is inhomogeneous on intermediate time scales or length scales,
{\it i.e.} the material forms 
a state consisting of a mixture of local
undoped antiferromganetic regions  and heavily-doped metallic regions.
While many mesoscopic motifs can result from this physics,
we have shown\cite{ute,spherical}
that the
typical consequence is an ordered ``stripe crystal'' or
fluctuating ``stripe liquid'' phase.  Within the stripe liquid phase,
there is also the possibility of a ``nematic stripe liquid'' phase in which
the stripes are positionally disordered, but are, on average, oriented
along a particular crystallographic direction, thus spontaneous breaking
the four-fold rotational symmetry of the crystal. 
The stripes are antiphase domain
walls in the antiferromagnetic order, and the doped holes are
concentrated in the domain walls, where the antiferromagnetic order vanishes.

\section{Concerning the microscopic origin of the stripes}

It is necessary to distinguish the various microscopic routes to stripe
phases, since they have rather different consequences\cite{LA,salkola,tranq}.
In one mechanism, stripes ({\it i.e.} long
period charge-density waves) arise from Fermi-surface nesting in a weakly
incommensurate system \cite{HFstripes}.  So far as is known,
in the Hartree-Fock approximation which captures the essence of the
Fermi surface instability, the phase separated state \cite{LA} and stripes 
\cite{HFstripes,LA} give the best variational solutions
of the two-dimensional Hubbard model at small values of doping.

The Hartree-Fock stripes
are, in fact, extremely close analogues of the solitons in polyacetylene, 
in that they are a lattice of antiphase domain walls whose 
mean separation is determined self-consistently from the hole
concentration so that the mid-gap bands are completely empty.  It is,
of course, a general feature of a Fermi-surface instability that the
order develops in such a way as to open gaps on the Fermi surface.
Moreover in the Hartree-Fock solution,
the instability is spin-driven, in the sense that there is a single
transition temperature below which the broken symmetry solution of the
Hartree-Fock equations is stable, and that the magnitude
of the spin-order parameter $m$
turns on with the usual mean-field exponent, $m \sim (T_c-T)^{1/2}$, while
the charge modulation turns on more weakly, ({\bf i.e.} $\sim m^2$).
In short, there are three clear features of stripes that arise from
a Fermi surface instability:  {\bf i)}  The transition is spin driven.
{\bf ii)}  In the low temperature phase, there are gaps (or pseudogaps) on 
the Fermi surface. {\bf iii)} The spacing
between domain walls is equal to $1/x$, where $x$ is the hole concentration.
(We have checked, by explicit calculation, that inclusion
of long-range Coulomb interactions in the Hartree-Fock caclulation has
little effect on this result \cite{salkola}.)
{\bf iv)} Although it is not easy to quantify, the notion of a Fermi-surface 
instability requires that the high
temperature phase should be a Fermi liquid, with quasiparticle scattering 
rates no greater than the ultimate low-temperature energy gap.

The situation is quite different if the stripes arise 
from  Coulomb-frustrated phase separation:  {\bf i)}  One
typically expects the transition to be charge-driven \cite{EKZ,tranq}.  
That is to say,
local spin order between the anti-phase domain walls can only develop
{\it after} the holes are expelled from these regions.  
In this case, general Landau-Ginzburg considerations\cite{EKZ},
lead one to expect either a first order transition, in which spin and charge
order turn on simultaneously, or  a sequence of transitions, in which
first the charge order and then the spin order appears as the temperature is
lowered.  {\bf ii)}  Since the stripe concentration is determined
primarily by the competition between the Coulomb interaction
and the local tendency to phase separation, we do not necessarily expect
that the spacing between stripes should be a simple function of $x$ and,
as a consequence, there
is no reason to expect the Fermi energy to lie in a gap or psuedogap;
in fact, the Fermi energy typically will lie in a region of
large density of states\cite{salkola}.  {\bf iii)} A high temperature Fermi 
liquid phase is not a prerequisite, since it is not an essential part of
the physics of frustrated phase separation. 

Although the formation of stripes is driven by frustrated phase
separation, their conformation is influenced by other more subtle effects.
{\bf i)}  As with any other charge-density wave system, commensurability
energies are likely to be important. For instance, in the 214 family of
superconductors (discussed below) 
either the stripe separation is nearly equal to 4 lattice 
spacings or the concentration of holes along a stripe is near to one per two 
sites. 
These two commensurabilities are compatible at $x=1/8$; otherwise the system
must choose which one (if either) to satisfy.  Commensurability effects
have been explored in some detail in doped La$_2$NiO$_4$ by Tranquada
and co-workers \cite{LNO}. Since systems of this sort can easily support
a disorder line\cite{pro}, commensurability effects can be important
even in a stripe liquid phase.  {\bf ii)}  The underlying crystalline
lattice possesses, at best, fourfold rotational symmetry, not full
rotational symmetry.  This gives rise to a lattice anisotropy energy which 
strongly affects the ordering of stripe phases\cite{spherical}, and
can also stabilize a nematic stripe-liquid phase.  
Moreover, the strength of the commensurability and anisotropy energies may 
be strongly influenced by structural phase transformations 
\cite{axe}.

\section{The La$_2$CuO$_4$ (214) family}

The ``high temperature superconductors'' are a group of materials which 
contain essentially equivalent CuO$_2$ planes separated by a variety of
interstitial charge reservoir regions which are the source of the doping.  
It is generally believed that
the basic physics is the same for all these materials, although
subtle differences sometimes make it difficult to distinguish 
the essential and universal from the 
inessential or material dependent.

We choose to focus on the so-called 214 sub-family of these materials. 
Not only are they among the best studied but neutron scattering structure 
factors provide direct ``photographic'' evidence of the topological character 
of the doping process.  In particular, we will contrast two slightly different 
materials of the family: 
La$_{2-x}$Sr$_x$CuO$_4$ and La$_{1.6-x}$Nd$_{0.4}$Sr$_x$CuO$_4$. In both cases,
the undoped system ($x=0$) is an antiferromagnetic insulator and the concentration
of doped holes in the CuO$_2$ planes is equal to the Sr concentration, $x$.
However, at the temperatures and dopant concentrations of interest to us here,
La$_{2-x}$Sr$_x$CuO$_4$ has an orthorhombic structure, whereas
La$_{1.6-x}$Nd$_{0.4}$Sr$_x$CuO$_4$ undergoes a structural phase transition
to a low temperature tetragonal (LTT) phase below about 70K.  Above 70K, the 
electronic properties
of the two materials are essentially indistinguishable for the same dopant
concentration $x$ but the LTT phase supresses
superconductivity at dopant concentrations $x < 20\%$ and, as has been recently
discovered\cite{tranq}, stabilizes an ordered ``striped'' state, 
in which dopant-induced
antiphase domain walls crystallize.

{\it The Undoped State:} The undoped state of all the high temperature superconductors
is a quasi two-dimensional antiferromagnetic insulator.   
At all temperatures below 1000K, there
are substantial antiferromagnetic correlations, although the system does not
finally order into a N{\'e}el state until a temperature of the order of 400K.  
Since the low-energy charge excitations involve the movement of charge from 
copper to oxygen, the material is known as a charge-transfer insulator. 
However, in common with a Mott insulator, the gap in the charge excitation
spectrum (roughly 2eV) is much greater than the scale of magnetic 
ordering \cite{sudip}.

{\it The Doped Insulator:}
A breakthrough in understanding the properties of the doped insulating state
was achieved recently through neutron scattering studies\cite{tranq} of a crystal of
La$_{1.6-x}$Nd$_{0.4}$Sr$_x$CuO$_4$ with $x=0.12$.  As the temperature is
lowered, there is a succession of transitions to 1) the LTT phase, 2) a 
charged-ordered state (charge-stripes) and 3) at a slightly lower temperature,
a period-doubling magnetically-ordered state.  This is a 
photograph of an ordered stripe phase of the sort predicted above.  
The factor of two between the periods of the charge and spin orders
implies that the charge is localized in magnetic antiphase domain walls, and
the fact that the charge order appears first implies that the physics is 
driven by charge ({\it i.e.} frustrated phase separation), rather than by 
Fermi surface or Hartree-Fock physics.  Moreover the
charge density along the stripes corresponds to a half-filled band (one doped 
hole per unit cell containing two Cu ions), which is consistent with the 
mobility required by kinetic phase separation and commensurability with the 
underlying lattice, but half as large as the prediction of Hartree-Fock 
theory.  Therefore, the origin of the stripe crystal phase of this material may
be attributed uniquely to frustrated phase separation.

La$_{1.6-x}$Nd$_{0.4}$Sr$_x$CuO$_4$ with $x=0.12$ is thought not to exhibit any
bulk superconductivity. In a manner that we believe we understand well 
\cite{EKZ}, the LTT phase stabilizes an ordered stripe phase and, 
the same time, supresses superconductivity.  That the ordered stripe phase is
in direct competition with superconductivity already suggests that it is an
extremely important piece of the physics.  However, there is more direct evidence.
La$_{1.6-x}$Nd$_{0.4}$Sr$_x$CuO$_4$ (with $x=0.12$) has static magnetic order
below about 50K, but
La$_{2-x}$Sr$_x$CuO$_4$ (with $x=0.15$) is magnetically disordered at all 
temperatures. Nevertheless the dynamical neutron scattering structure 
factors of both materials, measured  
at a frequency $\omega = $ 3 meV and T=40K \cite{ben}
(which effectively take a snapshot of the magnetic structures of the 
systems\cite{imtime}) 
are somewhat broadened versions of the elastic (Bragg)
scattering observed in La$_{1.6-x}$Nd$_{0.4}$Sr$_x$CuO$_4$.  Thus, there can
be no question that there exist substantial dynamic stripe fluctuations in
the optimally superconducting samples of La$_{2-x}$Sr$_x$CuO$_4$.  Moreover,
the fact that these fluctuations
appear at such low frequencies, implies that they are
very nearly frozen, even in the superconducting material!

This evidence is merely the latest and most graphic demonstration of 
topological doping in the high temperature superconductors.
It is a dramatic confirmation of our interpretation of other experiments,
as discussed in our earlier reviews, which includes 
analogues of most of the properties 
of polyacetylene, discussed above.  
A noteworthy property of the lightly-doped superconducting materials 
(known in the field as ``underdoped'') is the existence of a spin gap in the
normal state, which we have associated with local pairing correlations
without global phase coherence\cite{nature}.  Recent photoemission experiments\cite{shen}
appear to confirm this interpretation.  The only known ways of 
realizing this situation are in 
highly anisotropic systems ({\it e.g.} quasi one-dimensional where
the superconducting phase coherence is suppressed by thermal and quantum
fluctuations) or in  granular, spatially inhomogeneous, systems
in which global superconductivity is suppressed by
the small Josephson coupling between grains.  To some extent, both
of these features are present in the high temperature 
superconductors.

{\it The Overdoped Metal:}
It is conventional to classify the high temperature superconductors as 
``optimally doped'' ({\it i.e.} having the highest transition temperature in
a given class of materials), ``underdoped'' ({\it i.e.} having a lower dopant
concentration than optimal), and ``overdoped'' ({\it i.e.} having a larger
than optimal dopant concentration).  By now, the characteristics of the
members of these three classes are sufficiently well known that it is 
possible to make an assignment to a given material, even if systematic 
studies of the transition temperature as a function of dopant concentration 
are unavailable.
We have identified\cite{nature} 
``underdoped'' materials as doped insulators, in the sense of this article,
and have suggested that ``optimally doped'' is simply a crossover between 
underdoped and overdoped.
As in polyacetylene, the overdoped material appears to be more conventionally
metallic.  Certainly the conductivity at all temperatures
is much greater than $\sigma_{m}$, and it is much more isotropic
({\it i.e.} three dimensional) than in under and optimally doped materials,
although it still has a strange temperature dependence\cite{takagi}.  
Moreover, the
magnetic properties appear to be those of a conventional metal.
Thus, while it is not certain 
that the overdoped state can be classified as a Fermi liquid, it surely is a
more conventional metallic state than at lower doping.

The essential features of topological doping in the
214 family of high temperature superconductors are
summarized in Table II, which includes some additional properties   
discussed in our published papers.

\section{Other Cuprate Superconductors}
The existence of charge-density wave fluctuations adds a new dimension
to discussions of the physics of the {\hts} which, 
almost from the outset, have been
dominated by the interplay of antiferromagnetism and superconductivity.
It is clear that the three phenomena are simply different manifestations of 
a common underlying physical theme, and that the best way to investigate
any one of them is to find the material in which it shows up the most 
clearly. Nevertheless, although charge-density fluctuations are most
clearly manifested in the 214 family, it is 
natural to ask how their effects might be revealed in other materials.
Evidently it is desirable to have
a space-sensitive probe and, in the absence of suitable neutron or
X-ray scattering data, the obvious technique is angle-resolved photoemission 
spectroscopy (ARPES), which, until recently, was most reliable 
for Bi$_2$Sr$_2$CaCu$_2$O$_{8+x}$. We have undertaken a detailed study of
the implications of ordered and fluctuating charge stripes for the 
single-particle properties of {\hts}\cite{salkola}, and have found that it is
possible to understand the most striking features of the
ARPES data on Bi$_2$Sr$_2$CaCu$_2$O$_{8+x}$, especially the shape of the
Fermi surface and the existence of nearly dispersionless states in its
neighborhood. The essential ingredient in this interpretation of the data is 
a background of slowly-fluctuating stripes whose dynamics is determined by
collective effects (the competition between phase separation and the 
long-range Coulomb force), rather than the single-particle behavior.

Recently a detailed analysis of the ARPES experiments \cite{arpes}
has led to an understanding of the magnitude and temperature-dependence of
the width $\Gamma$ of the single-particle spectral functions, with two 
important consequences. First of all, in the normal state, $\Gamma$ is too 
large to be consistent with any kind of quasiparticle picture;  this 
observation strongly supports our analysis of the resistivity 
of the normal state \cite{bad}, which was based on the violation of the 
Ioffe-Regel condition \cite{IR}.
Secondly, it was found that $\Gamma$ decreases rapidly as the temperature
falls below T$_c$, indicating that quasiparticles are resurrected in the
superconducting state. In our picture, quasiparticle propagation is
inhibited in the normal state by the charge inhomogeneity, and especially
by the frustration of hole motion in the direction perpendicular to the
stripes, but phase order restores coherent particle motion, 
even though the stripes persist in the superconducting state. 

\section{What does this have to do with high temperature superconductivity?}
Our discussion has scarcely mentioned superconductivity which, after all,
provoked the worldwide interest in the oxide superconductors. But, as might 
be expected, an appreciation of the electronic structure and the 
physics that gives rise to it also leads to a natural mechanism of high 
temperature superconductivity. The fundamental driving force is the
frustration of the motion of holes in an antiferromagnet. This implies that
when the holes on the stripes tunnel into the intervening region, which has
a low charge density and strong antiferromagnetic correlations, they 
experience the very same effective attraction that would have caused them
to coalesce (phase separate) in the absence of the longer range Coulomb 
interaction\cite{bag}. Moreover, the Coulomb force, which so effectively inhibits 
pairing with a short coherence length and yet is widely ignored in discussions 
of the mechanism of high temperature superconductivity, has been 
incorporated into the picture from the outset. In fact, the holes
moving on a stripe form a quasi one-dimensional electron gas in an
``active'' environment with low-energy excitations in the spin (and possibly 
charge) degrees of freedom, which renormalize the kinetic energy of the
holes and mediate effective interactions between them \cite{EKZ}.
We have investigated this problem as a generalization of the standard theory 
of the one-dimensional electron gas, and have shown that it
provides a possible microscopic explanation of the phenomenon of
high temperature superconductivity \cite{EKZ}.
Whatever the mechanism, it must survive (and even thrive) on
significant charge-density wave (stripe) fluctuations, and account for their 
intimate competition with superconductivity. 
In this regard, we find the mechanism described above to be
an attractive possibility.  

Empirically and on general theoretical
grounds \cite{KE,nature}, it is clear that the optimal place for high temperature
superconductivity is in the crossover regime between the doped insulator and
overdoped regimes.  This suggests a general strategy for looking for superconductivity
in other doped correlated insulators\cite{KE}.

Acknowledgements: SK was supported in part by NSF grant \#DMR -93-12606.
This work also was supported by the Division of Materials Sciences,
U.S. Department of Energy, under contract DE-AC02-76CH00016. We have
enjoyed a continuing interaction with Dr. John Tranquada on the physics of 
oxide materials, and especially on charge ordering.

\newpage
\begin{table}
\begin{center}
\caption{Topological Doping in Polyacetylene (CH X$_y$)$_x$}
\begin{tabular}{||c||c||}
{\large \bf General Property} & {\large \bf Particular Manifestation}\\
\hline
\hline
{\bf Correlated insulator} &  {Commensurate (dimerized) CDW }\\
{ } & {Spontaneously broken reflection}\\
{ } & {and glide plane symmetry} \\
{ }  & {Charge gap} \\
{ } & {Spin gap} \\
{ } & {Single-particle gap} \\
\hline
\hline
{\bf Topological Doping} & { Soliton liquid or glass} \\
{ } & {Rapid restoration of symmetry}\\
{Doped insulator for $y < y_c$} & { $y_c \sim$ 1/ (soliton width)} \\
{ } & {Spin gap survives doping} \\
{ } & {Low-energy optical oscillator strength proportional to $x$}\\
{ } & {Charge gap feature persists with reduced oscillator strength}\\
{Inhomogeneous on intermediate} & 
{Transfer of oscillator strength to ``mid-gap'' states}\\
{  scales} & {Pinning of chemical potential in the gap}\\
{ } & {Conductivity $\ll \sigma_{m}$}\\
{ } & { }\\
{ } & { }\\
{ } & { }\\
\hline
\hline
{\bf ``Overdoped'' metallic state } & {3D metallic state}\\
{{\bf for  $y > y_c$} } & {Pauli spin susceptibility}\\
{ } & {3D Coherent quasiparticle propagation}\\
{ } & {Metallic magnitude of the conductivity}\\
{ } & {Density of states at E$_F$ consistent with band theory}\\
\end{tabular}
\end{center}
\end{table}
\begin{table}
\begin{center}
\caption{Topological Doping in La$_{2-x}$Sr$_x$CuO$_4$}
\begin{tabular}{||c||c||}
\hline
{\large \bf General Property} & {\large \bf Particular Manifestation}\\
\hline
\hline
{\bf Correlated insulator} &  {Commensurate (N{\'e}el) SDW }\\
{ } & {Spontaneously broken spin rotational} \\
{} & { \ \ \ \ and translational symmetry} \\
{ }  & {Charge gap} \\
{ } & {Spin waves} \\
{ } & {Single particle gap} \\
\hline
\hline
{\bf Topological doping} & { Stripe liquid, glass, or crystal} \\
{ } & {Rapid restoration of spin-rotational symmetry}\\
{Doped insulator for $x < x_c$} & { $x_c \sim$ 1/ (stripe width)} \\
{ } & {Spin gap appears at intermediate doping} \\
{ } & {Low-energy optical oscillator strength proportional to $x$}\\
{ } & {Charge gap feature persists with reduced oscillator strength}\\
{Inhomogeneous on intermediate}  & 
{Transfer of oscillator strength to ``mid-gap'' states}\\
{  scales} & {Pinning of chemical potential in the gap}\\
{ } & {Conductivity at high T $< \sigma_{m}$}\\
{ } & {Ordered striped phase in LTT structure}\\
{ } & {Macroscopic phase separation when dopants are mobile} \\
{ } & {Fluctuating stripes observed in neutron scattering}\\
\hline
\hline
{\bf ``Overdoped'' metallic state } & {3D metallic state}\\
{{\bf for $x > x_c$}} & {Destruction of spin gap}\\
{ } & {3D Coherent quasiparticle propagation}\\
{ } & {Metallic magnitude of the conductivity}\\
\hline
\end{tabular}
\end{center}
\end{table}


\begin{references}
\bibitem{elphonon}  For the present purposes, it is not extremely important
whether the interaction which produces the correlated insulating state is a 
direct electron-electron repulsion, as in the cuprate superconductors, 
or the electron-phonon interaction, as in polyacetylene.  We can always
view the system in a formalism in which we integrate out the phonons, and
consider the resulting induced interaction;  the only differences are that this
induced interaction is attractive and retarded.  Of course, for many purposes,
this distinction is important but, for illustrating the nature and
consequences of topological doping,  it is not.
\bibitem{currie}  In real samples there is always a small Curie-law susceptibility,
which is now thought to be an effect of the finite chain lengths and a corresponding
small concentration of neutral solitons in the ground state;  this type
of excitation would be absent in an ideal material.
\bibitem{symm}  In a purely one-dimensional system with no disorder,
the ground state symmetry would,
indeed, be restored at vanishingly small dopant concentration.  In the real material,
disorder and interchain interactions complicate the theory, and the experimental
situation is not completely resolved.  However, it is clear that even at very small
dopant concentrations, no static dimerization can be resolved in X-ray scattering.
\bibitem{imtime}  Technically, we are discussing imaginary time dynamics, 
as the structure factor is
a property of the thermal stationary state;  however,
thinking of the dynamic structure factor in this way is of considerable 
aid to intuition.
\end{references}
\end{document}